\documentclass{PoS}

\usepackage{graphicx}
\usepackage{epsfig}
\newcommand{\be}{\begin{equation}}
\newcommand{\ee}{\end{equation}}
\newcommand{\beq}{\begin{eqnarray}}
\newcommand{\eeq}{\end{eqnarray}}

\title{N to $\Delta$ transition form factors with $N_F=2+1$ domain wall 
fermions}

\ShortTitle{N to $\Delta$ transition form factors with $N_F = 2+1$ domain wall
fermions}

\author{C. Alexandrou\\
Department of Physics, University of Cyprus, CY-1678 Nicosia, Cyprus, and\\
Computation-based Science and Technology Research Center, The Cyprus Institute,
15 Kypranoros Str., 1645 Nicosia, Cyprus \\
E-mail: \email{alexand@ucy.ac.cy} }

\author{G. Koutsou\\
Forschungszentrum Juelich, D-52425 Juelich, Germany and \\
University of Wuppertal, Department of Physics, Wuppertal 42119, Germany\\
E-mail: \email{koutsou@ucy.ac.cy}}

\author{J. W. Negele\\
        Center for Theoretical Physics, Laboratory for
 Nuclear Science and Department of Physics, Massachusetts Institute of
Technology, Cambridge, Massachusetts 02139, U.S.A.\\
        E-mail: \email{negele@mit.edu}}

\author{\speaker{A. \'O Cais} and Y. Proestos\\
Computation-based Science and Technology Research Center, The Cyprus Institute,
15 Kypranoros Str., 1645 Nicosia, Cyprus \\
       E-mail: \email{a.ocais@cyi.ac.cy, y.proestos@cyi.ac.cy}} 

\author{A. Tsapalis\\
 Hellenic Naval Academy, Hatzikyriakou Ave., Pireaus 18539, Greece, and\\
Institute of Accelerating Systems and Applications, University of Athens,
Athens, Greece \\
        E-mail: \email{a.tsapalis@iasa.gr}}

\abstract{
The electromagnetic, axial and pseudoscalar nucleon to $\Delta$ form
factors are calculated using dynamical domain wall fermions 
at a lattice spacing of 
$a = 0.114$~fm on a lattice of spatial size $2.74$ fm and pion mass of 
$331$~MeV. Pion pole dominance and the Goldberger-Treiman relations
are examined.
}

\FullConference{The XXVII International Symposium on Lattice Field Theory\\
		 July 26-31 2009\\
		 Peking University, Beijing, China}

\begin{document}

\section{Introduction}

Form factors measured 
in electromagnetic and weak processes are fundamental probes of  hadron 
structure.  
Calculations of such observables using lattice QCD and, in particular,
 the nucleon form factors~\cite{nucleonFF, Syritsyn:2009mx, PRD_NN}
has intensified 
during the last couple of years due to 
improvements which allow full lattice QCD calculations
with controlled lattice systematics~\cite{Alexandrou:2009xk}.
The focus of the current work is the study of the electro-magnetic (EM) and
weak
N to $\Delta$ transition form factors (FFs).
Experiments on the N to  $\Delta$ EM transition have yielded
accurate results on the EM transition form factor for low momentum
transfer~\cite{exper} that point to deformation of the N/$\Delta$ system.
The axial N to $\Delta$ transition FFs
are experimentally not well known but there are ongoing
experiments using electroproduction of the $\Delta$ resonance
to measure 
the parity
violating asymmetry in N to $\Delta$.
Lattice QCD enables calculation of these fundamental quantities from first principle. 
Our previous calculation of these
form factors utilized quenched and dynamical Wilson as well as a 
hybrid scheme with domain wall (DWF) valence quarks on an improved staggered 
sea~\cite{PRD_GT,PRD_NDem,PRL2}. A study of the N to $\Delta$ transition
 using chiral dynamical quarks in a unitary approach 
is presented in this work where, in addition, we employ
 the coherent sink method~\cite{Syritsyn:2009mx}
in order to achieve the better statistical accuracy on the determination of the form factors.
\vspace{-0.2cm}

\section{Lattice Techniques}

\begin{figure*}[h]
%\vspace*{-0.3cm}
    \begin{minipage}[h]{7.5cm}
{\mbox{\includegraphics[height=7cm,width=8cm]{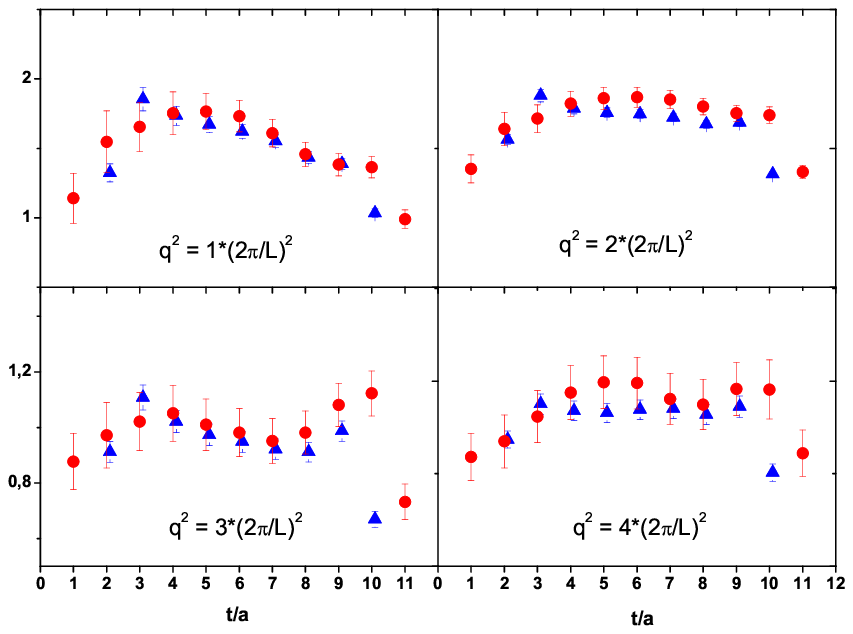}}}
    \caption{The ratio $S_1$ of Eq.~(3.2) versus $t/a$  
for a source-sink
separation 0.91~fm shifted by a time slice (blue triangles)
and 1.14~fm (red circles)
for the smaller non-zero $\vec{q}^2 $.
}
\label{fig:time-separation}
    \end{minipage}
    \hfill
 \begin{minipage}[h]{7.1cm}\hspace*{-0.4cm}
{\mbox{\includegraphics[height=7cm,width=7.5cm]{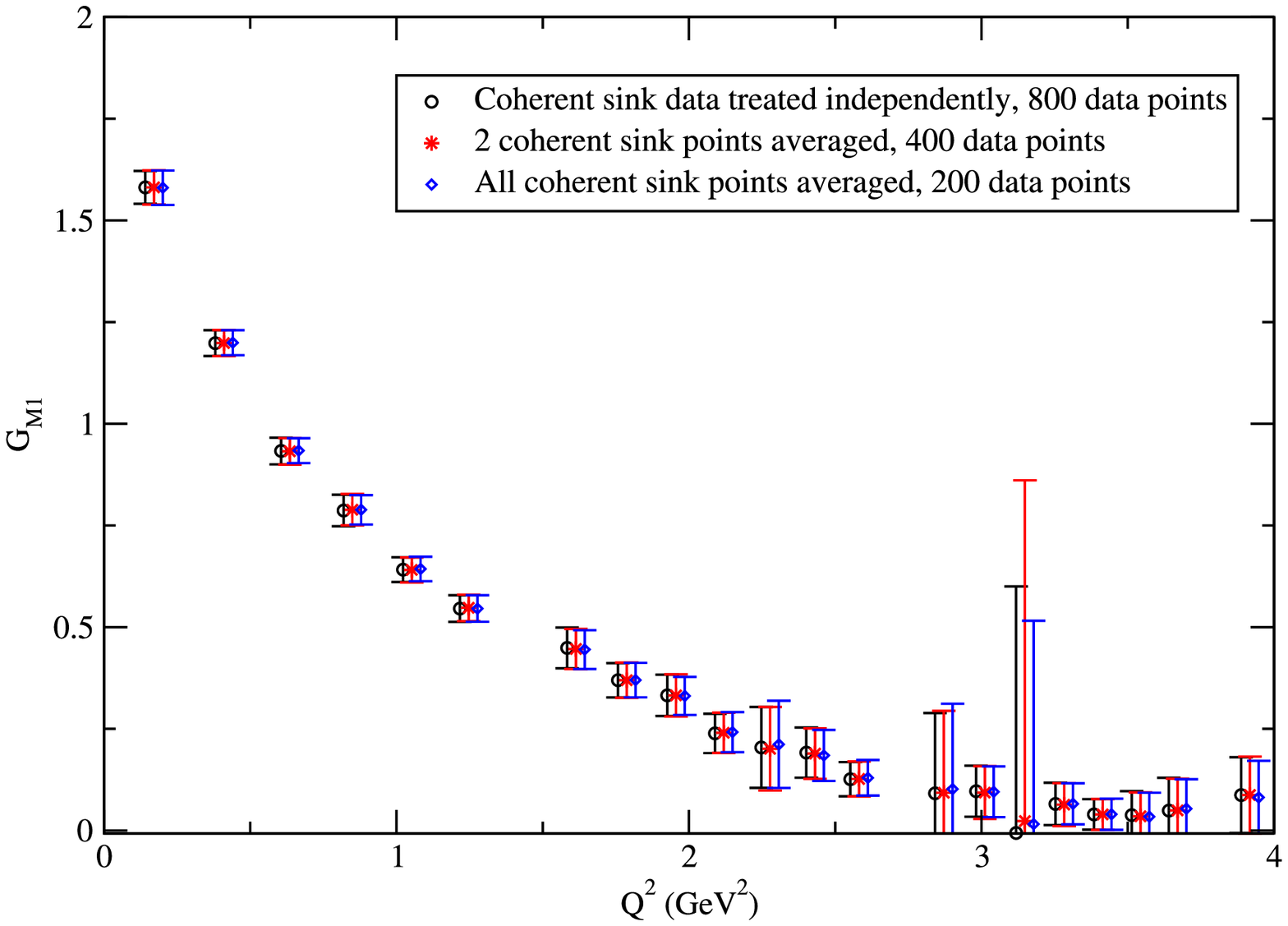}}}
    \caption{Dependence of the jackknife error for $G_{M1}(Q^2)$
on the coherent sink bin sizes. 
}
\label{fig:GM1sinks}
    \end{minipage}
\vspace*{-0.4cm}
\end{figure*}

We use $N_F = 2+1$ dynamical domain wall fermions
 generated by the  RBC and UKQCD collaborations~\cite{Allton:2008pn}. 
The lattice spacing  $a^{-1}= 1.73(3)$~GeV is fixed using the $\Omega^{-}$ mass.
The length of the fifth dimension is taken sufficiently large to suppress
chiral symmetry breaking. Fixing 
$L_5/a=16$ gives
an additive residual mass $\sim 10$\%  of the 
 light quark mass used in this work. 
%The mass of the pion and
%kaon are used to set the light and strange quark masses. The mass of
%the strange quark is over-estimated 
%by about $12\% $  as compared to  the physical value.
We consider configurations on a lattice of size $24^3\times 64$ 
corresponding to 
 pion mass of $0.331(1)$~GeV. We use the standard interpolating operators to
create nucleon and $\Delta$ states and employ  gauge invariant gaussian 
smearing of the quark fields with APE-smeared gauge fields optimized
for best suppression of excited states for the nucleon~\cite{PRD_NN}. 
 Suppressing
 excited state contributions in the three-point function
is particularly crucial since for this study a source-sink separation
of 0.9~fm is used. We show in Fig.~\ref{fig:time-separation} that
 extending the source-sink separation to 1.14~fm the plateau values for the dominant dipole form factor
$G_{M1}$, which are the most accurate, are consistent with a time-separation
of 0.9~fm, but with a two-fold increase in statistical errors.

The
three-point functions that are needed are given by 
\be 
\langle G_{\sigma}^{\Delta J_\mu N} (t_2, t_1 ;
{\bf
p}^{\;\prime}, {\bf p}; \Gamma_\tau) \rangle  = 
\sum_{{\bf x}_2, \;{\bf
x}_1} e^{-i {\bf p}^{\prime} \cdot {\bf x}_2 } e^{+i {\bf q}
\cdot {\bf x}_1 } \; \Gamma_\tau^{\beta \alpha}
\langle \Omega | 
T\left[\chi_{\Delta}^{\sigma \alpha} ({\bf x}_2,t_2) J_\mu({\bf
x}_1,t_1) \bar{\chi}_N^{\beta} ({\bf 0},0) \right] |\Omega
\rangle  
\label{3pt} 
\ee
where $J_\mu(x)$ is a local current, ${\bf q} = {\bf p}^{\prime}-{\bf p}$
is the momentum transfer,  $\sigma$  is the Lorentz vector index for the $\Delta$
and
$\Gamma_{\tau}$ projection matrices in Dirac space~\cite{PRD_NDem}.
The large Euclidean time limit of the ratio
\small 
\be 
%\vspace*{-0.8cm} 
R^{J}_\sigma (t_2, t_1; {\bf p}^{\; \prime}, {\bf p}\; ; \Gamma_\tau ; \mu)
=   \frac{\langle G^{\Delta J_\mu N}_{\sigma} (t_2,
t_1 ; {\bf p}^{\;\prime}, {\bf p};\Gamma ) \rangle \;} {\langle
G^{\Delta \Delta}_{ii} (t_2, {\bf p}^{\;\prime};\Gamma_4 ) \rangle
\;}  
\biggr [\frac{\langle G^{\Delta \Delta}_{ii} (t_2, {\bf
p}^{\;\prime};\Gamma_4 ) \rangle}{ \langle
G^{N N} (t_2, {\bf p};\Gamma_4 ) \rangle }\> 
\frac{ \langle G^{N N}(t_2-t_1, {\bf
p};\Gamma_4 ) \rangle \;\langle G^{\Delta \Delta}_{ii} (t_1, {\bf
p}^{\;\prime};\Gamma_4 ) \rangle} {\langle G^{\Delta \Delta}_{ii}
(t_2-t_1, {\bf p}^{\;\prime};\Gamma_4 ) \rangle \;\langle
G^{N N} (t_1, {\bf p};\Gamma_4 ) \rangle} \biggr ]^{1/2} 
\label{ratio}
\ee 
%&\> & \stackrel{t_2 -t_1 \gg 1, \;\; t_1 \gg
%1}{\Rightarrow} \Pi^J_{\sigma}({\bf p}^{\; \prime}, {\bf p}\; ;
%\Gamma_\tau ; \mu) \;  
%\eeq
\normalsize
yields a time-independent function 
$\Pi_{\sigma}({\bf p}^{\; \prime}, {\bf p}\; ;\Gamma_\tau ; \mu) \;$ (plateau region).  In addition, all field renormalization
constants  cancel and therefore $\Pi_\sigma$ is a combination 
of the Lorentz invariant form factors  and known kinematical factors. 
We use sequential inversions through the sink to evaluate the three-point function of Eq.~(\ref{3pt}). In this method the quantum numbers of the hadron
 are fixed, which means that a particular
value of $\sigma$ and $\Gamma_\tau$ must be chosen.  
This freedom 
 is exploited in the construction of  sources for the sequential propagator
with the goal to produce  optimal linear 
combinations of $\Pi_\sigma$
involving a maximal set of  momentum vectors, thereby obtaining
 a maximum number of statistically independent measurements~\cite{PRD_GT}.
It turns out that three such sinks suffice for
achieving this goal and enable us to extract  the momentum dependence of 
the electromagnetic, axial and 
pseudoscalar N to $\Delta$ FFs accurately. 
A new ingredient of the current work is the use of the {\it coherent sink 
technique}~\cite{Syritsyn:2009mx}  in order to reduce the statistical noise. This
 consists of creating four sets of forward propagators for each configuration 
by placing sources at:\\
$(\vec{0},0), \quad (\vec{L}/2,16), \quad (\vec{0},32)\quad {\rm and} 
\quad (\vec{L}/2,48).$ \\
From each source $(\vec{x}_i,T_i)$,
a zero-momentum projected $\Delta$ source is constructed at $T_0$ away, i.e. at
$(\vec{x}_i,T_i+T_o)$ and a single coherent backward propagator is calculated
in the
simultaneous presence of all four sources. The cross terms
that arise vanish by gauge invariance when averaged over the ensemble.
The forward propagators 
are already computed by the LHPC Collaboration~\cite{Syritsyn:2009mx} and therefore
 we effectively obtain
 four measurements at the cost of one. This assumes large enough
time-separation
between the four sources to suppress contamination among them. An open question
is whether there exists correlation among these four measurements.
In
Fig.~\ref{fig:GM1sinks} we show the dependence of the jackknife error on 
$G_{M1}$ for different coherent sink bin sizes, which verifies that 
cross-correlations between the different sinks are absent.

The full set of data obtained at a given $Q^2$ value is analyzed simultaneously
by a global $\chi^2$ minimization using the singular value decomposition of an
overconstrained linear system~\cite{PRD_GT}. All the results presented here 
are obtained by analyzing $200$ configurations or a total
of  $200 \times 4 = 800$   measurements of the ratio given in Eq.~(\ref{ratio}).
\noindent

\section{Electromagnetic N to $\Delta$ Transition form factors}

The electromagnetic transition matrix element is decomposed in terms of 
three Sachs (FFs)
\be
  \langle\Delta(p',s')\vert j_\mu \vert N(p,s)\rangle =
  i \,\sqrt{\frac{2}{3}} \; \biggl(\frac{m_{\Delta}\; m_N}
{E_{\Delta}({\bf p}^\prime)\;E_N({\bf p})}\biggr)^{1/2} \bar{u}_\sigma (p',s')
  {\cal{O}}_{\sigma\mu} u(p,s) 
\ee
with
$$  {\cal O}_{\sigma\mu} = G_{M1} (q^2) K_{\sigma\mu}^{M1} + G_{E2}(q^2)
  K_{\sigma\mu}^{E2} + G_{C2} (q^2) K_{\sigma\mu}^{C2} 
$$
where
$K_{\sigma\mu}^{M1}, K_{\sigma\mu}^{E2}$ and 
$K_{\sigma\mu}^{C2}$
are known  kinematical factors~\cite{PRD_NDem}.
In this work we present results for the dominant magnetic dipole 
form factor $G_{M1}(q^2)$. 
Following Ref.~\cite{PRD_NDem} we construct the
optimized three-point function $S_1$ from which $G_{M1}(Q^2)$ is directly 
determined
\be
\label{S1} 
S_1({\bf q};\mu)= \sum_{\sigma=1}^3\Pi_\sigma({\bf 0},-{\bf
q}\; ;
\Gamma_4 ;\mu) =  i A \biggl\{ (p_2-p_3)\delta_{1,\mu} 
 + (p_3-p_1)\delta_{2,\mu} + (p_1-p_2)\delta_{3,\mu} \biggr\}
G_{M1}(Q^2) 
\quad.
\ee
\noindent
\begin{figure}[h]
\vspace*{-1.2cm}
 \begin{minipage}[h]{7.5cm}
{\mbox{\includegraphics[height=6.5cm,width=7.5cm]{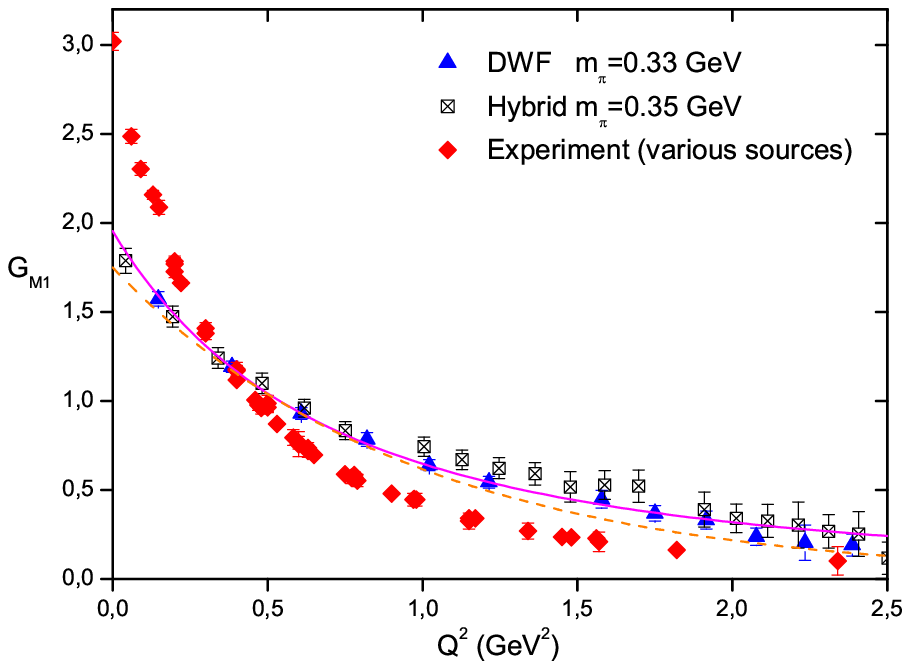}}}
 \vspace*{-0.5cm}
    \caption{$G_{M1}(Q^2)$ using DWF 
fermions and using the hybrid action. 
The diamonds show experimental results.
The solid (dashed) line is a fit to 
dipole (exponential) form for the DWF data.
}
\label{fig:GM1fit} 
    \end{minipage}
    \hfill
    \begin{minipage}[h]{7.1cm}\hspace*{-0.4cm}
{\mbox{\includegraphics[height=6.5cm,width=7.5cm]{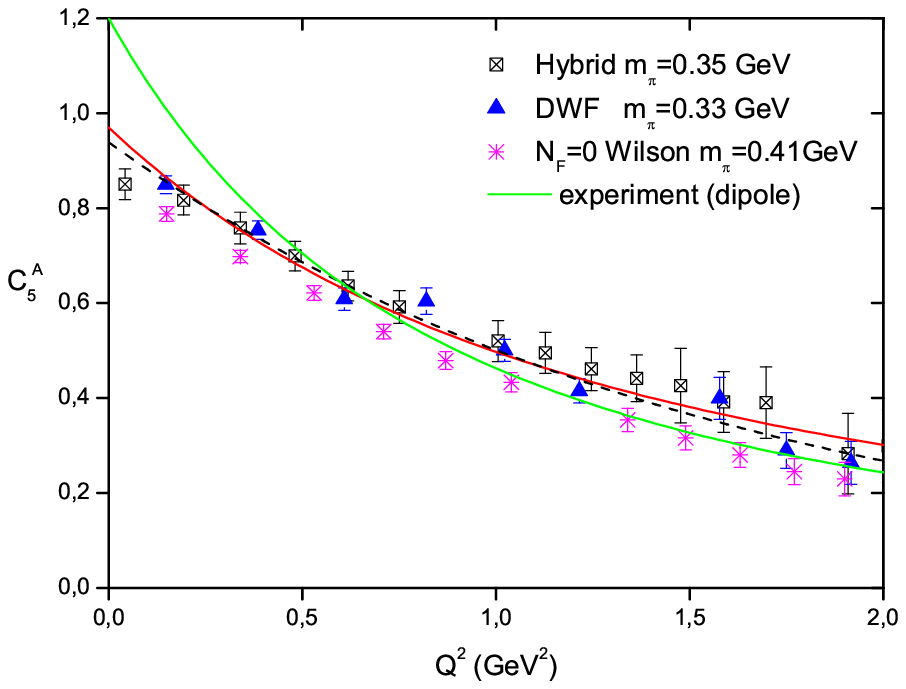}}}
 \vspace*{-0.5cm}
   \caption{$C_5^A$ 
for  DWF, the hybrid action and  quenched Wilson fermions (  
$m_\pi$ = 410 MeV)~\cite{PRD_GT}. The green line is a dipole fit
to experimental data~\cite{Kitagaki}. 
The solid (dotted) line is a fit to dipole (exponential) form of the DWF data.
}
\label{fig:CA5}
    \end{minipage}
\vspace*{-0.2cm}
\end{figure}

In Fig.~\ref{fig:GM1fit} we show the results of this work
on  $G_{M1}(Q^2)$ using DWF. These are
compared with previous results obtained with a hybrid action that uses
Asqtad improved staggered fermions generated by the MILC collaboration and
domain wall valence quarks~\cite{PRD_NDem}. 
The pion mass in the DWF calculation is 331 MeV
and in the hybrid action 350 MeV. These values are close enough to
allow a direct comparison. Indeed the results are in very good agreement.
Fits to a dipole form,  $g_0/(1+Q^2/m_0^2)^2$, as well as to an 
exponential form 
$\tilde{g_0} 
\;exp(-Q^2/\tilde{m_0}^2)$  described equally well the lattice results.
%We obtain similar values for the mass-parameters:
% $m_0 = 1.164(20)$~GeV and  $\tilde{m_0}=1.021(17)$~GeV.
 A compilation of the experimentally available data (for more
details see Ref.~\cite{PRD_NDem}) is also shown in Fig.~\ref{fig:GM1fit}
showing a clear disagreement between lattice results and experiment.
This is reflected in the value of the  dipole mass of $m_0 = 0.78$ GeV
obtained by performing a dipole form fit to the experimental data
as compared to  $m_0 = 1.164(20)$~GeV for the lattice results.
A possible explanation for the faster falloff of the experimental data  
maybe the lack of significant
chiral quark effects --or equally the lack of strong pion cloud-- from the
still heavy pion mass ensembles that are utilized. Similar 
behavior is also observed for the nucleon electromagnetic 
form factors~\cite{nucleonFF}, that may again point to the
importance of chiral quark effects. The N to $\Delta$
 case is particularly clean since there is
no ambiguity regarding disconnected contributions and thus  the flatter 
dependence observed in the N to $\Delta$ EM FFs must be
 of different origin.
The large disagreement observed here, however, would require large pion cloud
effects to set in as we lower the pion mass. Such large pion effects have been
shown to arise in chiral expansions~\cite{Pascalutsa}
 and it is thus  interesting to perform 
the calculation for $m_\pi<250$~MeV where they are expected to set in.  
We are currently analyzing results to extract the subdominant FFs,  $G_{E2}$ and $G_{C2}$ using the same DWF configurations.

\section{Electroweak N to $\Delta$ Transition form factors and 
Goldberger-Treiman relations}

We consider nucleon to $\Delta$ matrix elements of the axial and
pseudoscalar currents defined by
\be
A_{\mu}^a(x)= \bar{\psi}(x)\gamma_\mu \gamma_5\frac{\tau^a}{2}\psi(x) 
\hspace{1cm} , \hspace{1cm}
P^a(x)= \bar{\psi}(x)\gamma_5 \frac{\tau^a}{2}\psi(x) 
\label{currents}
\ee
where $\tau^a$ are the three Pauli-matrices acting in flavor space
and $\psi$ the  isospin doublet quark field.
The invariant proton to $\Delta^+$ weak matrix element is expressed
in terms of four transition
form factors in the Adler representation as
\beq
<\Delta(p^{\prime},s^\prime)|A^3_{\mu}|N(p,s)> &=& i\sqrt{\frac{2}{3}} 
\left(\frac{m_\Delta m_N}{E_\Delta({\bf p}^\prime) E_N({\bf p})}\right)^{1/2}
\bar{u}_{\Delta^+}^\lambda(p^\prime,s^\prime)\nonumber \\
&\>&\hspace*{-5cm}
\biggl[\left (\frac{C^A_3(q^2)}{m_N}\gamma^\nu + \frac{C^A_4(q^2)}{m^2_N}p{^{\prime \nu}}\right)  
\left(g_{\lambda\mu}g_{\rho\nu}-g_{\lambda\rho}g_{\mu\nu}\right)q^\rho
+C^A_5(q^2) g_{\lambda\mu} +\frac{C^A_6(q^2)}{m^2_N} q_\lambda q_\mu \biggr]
u_P(p,s).
\label{NDaxial}
\eeq
The form factors
$C^A_3(q^2)$ and $C^A_4(q^2)$ belong to the transverse part of the axial 
current and are both suppressed~\cite{PRL2} relative to the dominant 
form factors $C^A_5(q^2)$ and   $C^A_6(q^2)$.  The latter two are the
equivalent to the nucleon axial FFs $G_A(Q^2)$ and $G_p(Q^2)$ respectively~\cite{PRD_GT}.

The pseudoscalar transition form factor 
$G_{\pi N\Delta}(q^2)$, is defined via
\small
\be 
 2m_q<\Delta(p^\prime,s^\prime)|P^3|N(p,s)> = i\sqrt{\frac{2}{3}}
\left(\frac{m_\Delta m_N}{E_\Delta({\bf p}^\prime) E_N({\bf p})}\right)^{1/2}
\frac{f_\pi m_\pi^2 \>G_{\pi N\Delta}(q^2)}
{m_\pi^2-q^2}
\bar{u}_{\Delta^+}^\nu(p^\prime,s^\prime)\frac{q_\nu}{2m_N} u_P(p,s)
\label{gpiND} \quad.
\ee
\normalsize
Taking matrix elements of the axial Ward-Takahashi identity
$
 \partial^\mu A_\mu^a=2m_qP^a
$
leads to the non-diagonal Goldberger-Treiman (GT) relation
\beq
 C_5^A(q^2)+\frac{q^2}{m_N^2} C_6^A(q^2) = 
\frac{1}{2m_N}\frac{G_{\pi N \Delta}(q^2)f_\pi m_\pi^2}{m_\pi^2-q^2} \quad.
\label{GTR_ND}
\eeq
%The normalization of the matrix element (\ref{gpiND}) can be understood if
%we also consider 
The PCAC relation on the hadronic level 
$
\partial^\mu A_\mu^a=f_\pi m_\pi^2 \pi^a,
$
relates the pseudoscalar current to the pion field operator and 
therefore provides the connection to the 
phenomenological $\pi N\Delta$ strong coupling  
$g_{\pi N\Delta}=G_{\pi N\Delta}(0)$ that appears in Eq.~(\ref{GTR_ND}).
Assuming pion pole dominance we can
relate the form factor $C_6^A$ to $G_{\pi N\Delta}$
via:
\beq 
\frac{1}{m_N}C_6^A(q^2)&\sim&\frac{1}{2}\frac{G_{\pi N\Delta}(q^2) f_\pi}
{m_\pi^2-q^2}
\label{GP}
\eeq
Substituting in Eq.~(\ref{GTR_ND}) we obtain the simplified
Goldberger-Treiman relation
\beq
G_{\pi N \Delta}(q^2)\>f_\pi &=& 2m_N C_5^A(q^2)
\label{GTR}
\eeq
in complete analogy to the well known GT relation which holds in the
nucleon sector.
Pion pole dominance therefore fixes completely the ratio 
$C_6^A(q^2)/C_5^A(q^2)$ as a pure monopole term
\be
\frac{C_6^A(q^2)}{C_5^A(q^2)} = \frac{m_N^2}{m_{\pi}^2 -q^2}
\quad.
\label{monopole}
\ee
The goal here  is to calculate $C_5^A(q^2)$, $C_6^A(Q^2)$ and 
$G_{\pi N \Delta}(Q^2)$
and check the GT relations using dynamical DWF.
The relevant three-point functions required for
the calculation of these FFs are 
obtained at a {\it minimal} extra cost using the
sequential propagators produced from the optimized
nucleon to $\Delta$ source $S_1$ and in addition $S_2$ which is
also used  
for the electromagnetic transition study of the subdominant FFs.
The detailed expressions are given in Ref.~\cite{PRD_GT}.
%These optimal sources in the axial sector lead to the relations
%\small
%\beq 
% S_1^A(\mathbf{q};j) = \sum^3_{\sigma=1}\Pi^A_\sigma({\bf 0},-\mathbf{q};\Gamma_4;j) 
%= i B
%\Bigg[-\frac{C^A_3}{2}\bigg\lbrace (E_N-2m_\Delta+m_N)+
%\left(\sum_{k=1}^3 p^k\right)\frac{p^j}{E_N+m_N} \bigg\rbrace 
%\nn
%-\frac{m_\Delta}{m_N}(E_N-m_\Delta)C^A_4
%+m_N C^A_5-\frac{C^A_6}{m_N}p^j\left(\sum_{k=1}^3 p^k\right)\Bigg] ,
%\label{S1A}
%\eeq
%\beq
%S_1^A(\mathbf{q};4) = \sum^3_{\sigma=1}\Pi^A_\sigma({\bf 0},-\mathbf{q};\Gamma_4;4) 
%= B
% \sum_{k=1}^3 p^k\Bigg[C^A_3+\frac{m_\Delta}{m_N}C^A_4
%+\frac{E_N-m_\Delta}{m_N}C^A_6\Bigg]
%\label{S1_4}
%\eeq
%\beq
%\vspace{-0.3cm}
%S_2^A(\mathbf{q};j)= \sum^3_{\sigma\ne k=1}\Pi^A_\sigma({\bf 0},-\mathbf{q};\Gamma_k;j)= 
%i \frac{3 A}{2}\Bigg[\left(\sum_{k=1}^3 p^k\right) 
%\left(\delta_{j,1}(p^2-p^3)+
%\delta_{j,2}(p^3-p^1)+\delta_{j,3}(p^1-p^2)\right)C^A_3\Bigg], \nn 
%\label{S2}
%\eeq
%\normalsize
%where  $\;\;  A=\frac{B}{(E_N+m_N)}, \quad
%B=\sqrt{\frac{2}{3}}\frac{\sqrt{\left(E_N+m_N\right)/E_N}}{3m_N} .
%$
\begin{figure}[h]
    \begin{minipage}[h]{7.5cm}\hspace*{-0.4cm}
{\mbox{\includegraphics[height=6.5cm,width=7.5cm]{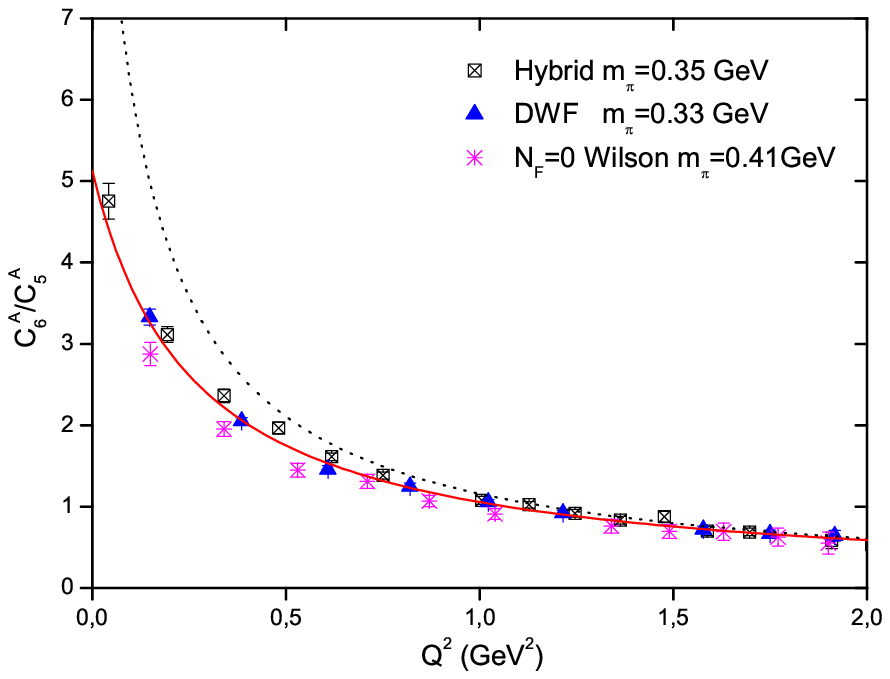}}}
\vspace*{-0.3cm}
    \caption{The ratio $C_6^A/C_5^A$ versus $Q^2$. The dotted line refers
to the DWF results and is the pion pole dominance prediction of Eq.~(4.5).
The solid line is a fit to a monopole form. 
}
\label{fig:CA6ovCA5}
    \end{minipage}
\hfill
    \begin{minipage}[h]{7.1cm}\hspace*{-0.4cm}
{\mbox{\includegraphics[height=6.5cm,width=7.5cm]{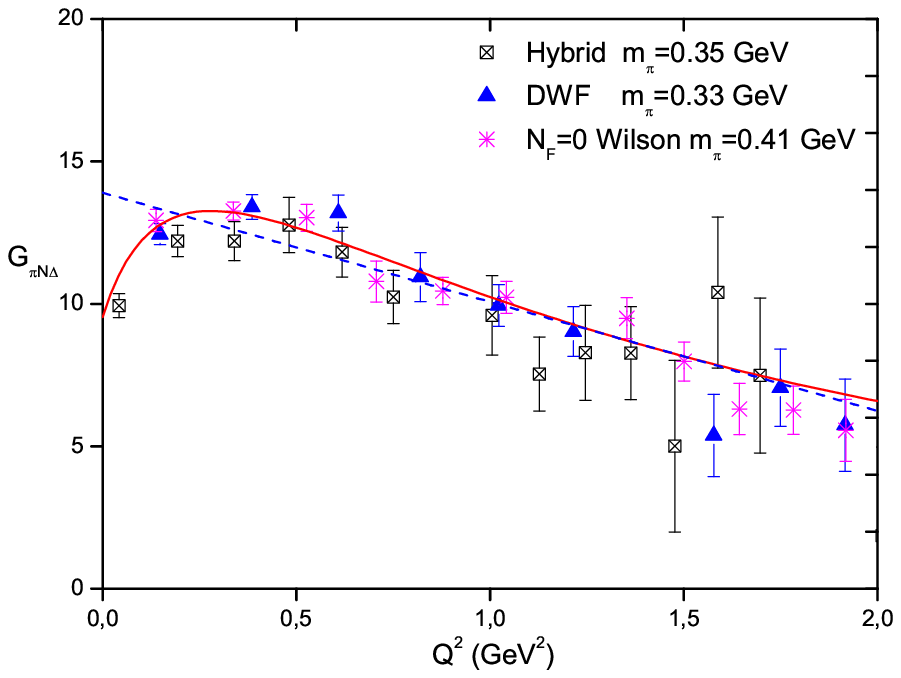}}}
 \vspace*{-0.3cm}
   \caption{$Q^2$-dependence of the pseudoscalar transition form factor
$G_{\pi N \Delta}$. The solid line is a fit to pion pole dominance form of 
Eq.~(4.9). The dashed line is a linear fit. The strong coupling
constant $g_{\pi N \Delta}$ is the value at $Q^2 = 0$.
}
\label{fig:GPND}
    \end{minipage}
\vspace*{-0.2cm}
\end{figure}

In Fig.~\ref{fig:CA5} we compare our results for $C_5^A$ using
DWF to those obtained previously using the hybrid action and quenched
 Wilson fermions at
similar pion masses~\cite{PRD_GT,PRL2}.
 The $Q^2$ dependence is well described by a dipole
Ansatz
%$
%\frac{g_0}{(Q^2/m_A^2+1)^2}
%$
yielding  $C_5^A(0)= 0.970(30)$ and a dipole mass $m_A = 1.588 (67)$~GeV.
This is to
be compared with the value
 $m_A = 1.28 \pm 0.10$~GeV extracted by a dipole fit to the
 available experimental 
data~\cite{Kitagaki}. 
As in the case of $G_{M1}(Q^2)$, we observe a flatter slope for 
the lattice data, reflected in the larger value of the axial mass $m_A$
extracted for the lattice results.
% In addition we show with the dotted line a fit 
%to an exponential form  $\tilde{g}_0 e^{-Q^2/\tilde{m}_A^2}$ which is almost
%indistinguishable from the dipole, for a mass $\tilde{m}_A = 1.262(36)$~GeV.
%Notice that the $C_5^A$ data for the hybrid MILC lattice in the 
%$Q^2 < 0.5$~GeV$^2$ regime are corrected~\cite{erratum} with respect to the
%values appearing in~\cite{PRD_GT}. A smaller correction to the lowest 
%$C_6^A(Q^2)$
%point for the hybrid action was also noted~\cite{erratum}.
%The fitted dipole mass for the hybrid
%$C_5^A$ is $m_A = 1.795(38)$~GeV, as compared to $1.534(36)$ for the 
%quenched Wilson volume.

In Fig.~\ref{fig:CA6ovCA5} we show the
ratio $C_6^A/C_5^A$. The dotted line shows the pion pole dominance
prediction of Eq.~(\ref{monopole}) where for  $m_N$ and $m_\pi$ we use
the lattice 
values calculated for  DWF. The predicted curve does not  describe 
the data at low $Q^2$ i.e. in the regime where strong pion cloud effects
are expected.
Fitting to a monopole form
$ c_0/(Q^2/m^2+1) $
describes satisfactorily the ratio yielding a heavier mass parameter $m$
 than the lattice value of the pion mass.
This behavior has been observed also for the other actions~\cite{PRD_GT}.

The pseudoscalar form factor $G_{\pi N \Delta}(q^2)$ is 
determined
optimally from the source $S_1$ with a pseudoscalar current 
operator insertion:
\small
\beq  
 S^P_1({\bf q}\; ;\; \gamma_5) = \sum_{\sigma=1}^3
\Pi_\sigma^P ({\bf 0},-{\bf q}\; ;
\Gamma_4 ;\gamma_5) = \sqrt{\frac{2}{3}}\sqrt{\frac{E_N+m_N}{E_N}} 
   \left[\frac{q_1 + q_2 + q_3}{6 m_N} \frac{f_\pi m_\pi^2} 
{2 m_q (m_\pi^2 + Q^2)} \right]\; G_{\pi N \Delta} (Q^2)
\label{PND optimal}
\eeq 
\normalsize
We use
the value $f_\pi = 0.1052(7)$~GeV for the pseudoscalar pion decay constant determined in Ref.~\cite{Allton:2008pn}.
 The quark mass
$m_q$  is calculated 
through the Axial Ward Identity by constructing 
 a suitable ratio of local-smeared and
smeared-smeared two-point functions of the axial and pseudoscalar 
currents~\cite{PRD_GT}. This requires only knowledge of the axial current
renormalization $Z_A$, which is determined to be $Z_A = 0.7197(9)$
(Yamazaki {\it et al} in \cite{nucleonFF}), where also $Z_V = Z_A$ holds
up to a small $O(a^2)$ error for a chiral action~\cite{Allton:2008pn}.

In Fig.~\ref{fig:GPND} we compare results on
 $G_{\pi N \Delta}(q^2)$ using dynamical DWF to those
obtained with the hybrid action and in the quenched theory~\cite{PRD_GT}.
The solid line is a one-parameter fit to the form
\be
G_{\pi N \Delta}(Q^2)=K \>\frac{Q^2/m_\pi^2+1}{(Q^2/m_A^2+1)^2(Q^2/m^2+1)} 
\label{fit G}
\ee
expected
 if the validity of Eq.~(\ref{monopole}) is assumed. 
The fit-parameter $K$ provides an estimate of the strong coupling
$g_{\pi N \Delta} = G_{\pi N \Delta} (0) = 9.6(2)$. A straight line 
fit of the form
$
G_{\pi N\Delta}(Q^2) \sim \biggl(1 -\Delta \frac{Q^2}{m_\pi^2}\biggr)
$
as shown by the dashed line, would lead to an estimate 
$g_{\pi N \Delta} = 13.9(6)$. Thus a reliable evaluation of
$g_{\pi N \Delta}$ requires further understanding of the behavior
at low $Q^2$  and in particular of the decrease observed in the
hybrid action at $Q^2$ close to zero.

\begin{figure}[h]
\vspace*{-0.3cm}
 \begin{minipage}[h]{7.5cm}
{\mbox{\includegraphics[height=6.5cm,width=7.5cm]{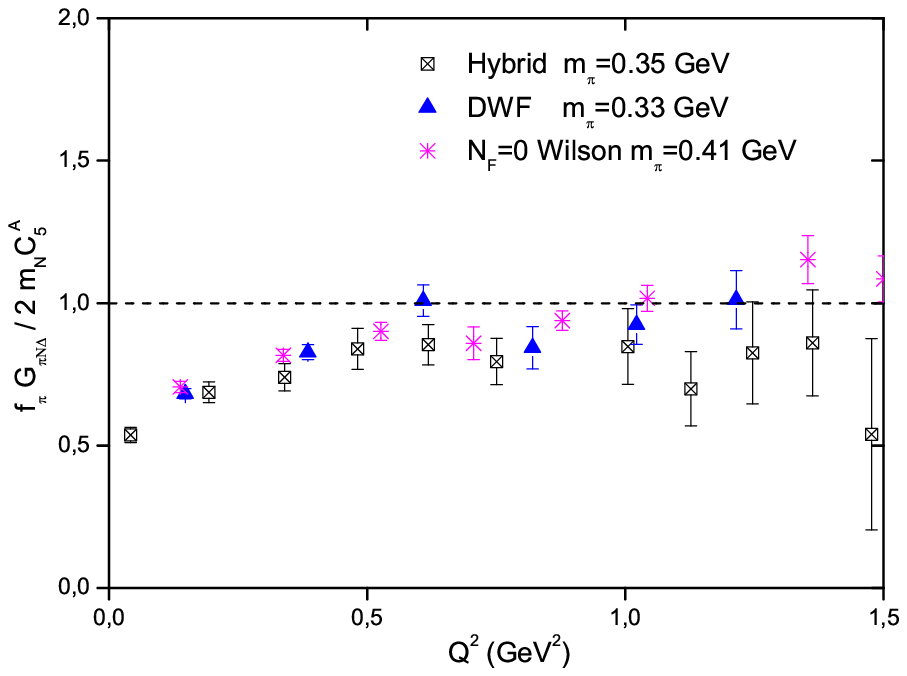}}}
\vspace*{-0.3cm}
    \caption{The ratio $f_\pi G_{\pi N\Delta}(Q^2)/m_N C_5^A(Q^2)$ as a function of $Q^2$ relating to the GT validity. 
}
\label{fig:GTR1} 
    \end{minipage}
    \hfill
    \begin{minipage}[h]{7.1cm}\hspace*{-0.4cm}
{\mbox{\includegraphics[height=6.5cm,width=7.5cm]{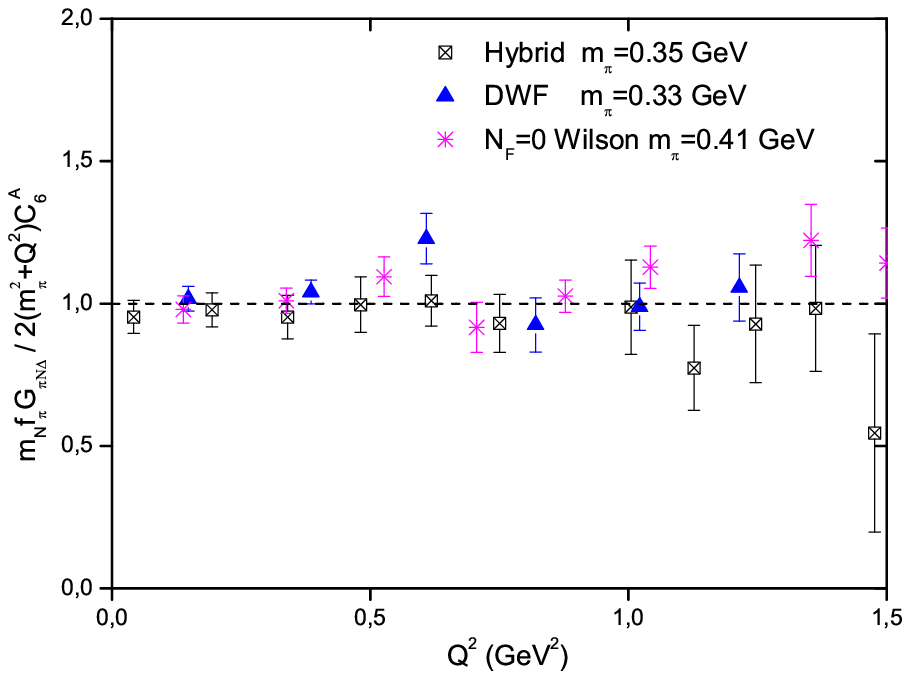}}}
\vspace*{-0.3cm} 
   \caption{The ratio $m_N f_\pi G_{\pi N\Delta}(Q^2) / 2(m_{\pi}^2 + Q^2)C_6^A(Q^2)$ that relates to the validity of Eq.~(4.5).  
}
\label{fig:GTR2}
    \end{minipage}
\vspace*{-0.2cm}
\end{figure}

In Fig.~\ref{fig:GTR1} we show the ratio  
$f_\pi G_{\pi N\Delta}(Q^2)/m_N C_5^A(Q^2)$,
which should be unity if  the non-diagonal GT relation of
Eq.~(\ref{GTR}) is satisfied.
 Deviations from this relation are evident in the low $Q^2$ regime
and they are present
for all actions to the same degree which is surprising since
one might have expected a better behaviour for DWF.
At higher momentum transfers ($Q^2 > 0.5$~GeV$^2$)  the relation is satisfied 
for all actions.
On the other hand, the relation given in Eq.~(\ref{monopole}) that assumes pion pole
dominance to relate $C_6^A$ to $C_5^A$ is satisfied excellently by the lattice
data for all three actions.
This agreement is shown in Fig.~\ref{fig:GTR2} where the ratio 
$m_N f_\pi G_{\pi N\Delta}(Q^2) / 2(m_{\pi}^2 + Q^2)C_6^A(Q^2)$ 
is everywhere consistent with unity.

\section{Summary and Conclusions}

The nucleon to $\Delta$ electromagnetic, axial and pseudoscalar transition 
form factors are calculated using $N_f=2+1$ dynamical domain wall fermions
 for pion mass
of 0.33 GeV. The dominant form factors $G_{M1}$ and $C_5^A$ show
slower falloff with $Q^2$ as compared to experiment. A possible
explanation maybe  that the pion cloud is still not
fully developed, at pion mass of 0.33~GeV. 
We examined the Goldberger-Treiman relations and found
that they are satisfied for $Q^2>0.5$~GeV$^2$ as was previously observed
for Wilson fermions and when using a hybrid action. 
Pion pole dominance relating the axial form factor $C_6^A$ and
the pseudoscalar form factor $G_{\pi N \Delta}$  
is satisfied for all values of $Q^2$ irrespective of the 
lattice action used.
Extraction of the strong coupling constant 
$g_{\pi N \Delta}$ requires special care since
 we need a better understanding of the low $Q^2$ behavior of
the pseudoscalar matrix element. A calculation on a finer lattice
using domain wall fermions
is underway to check for any cut-off effects 
as well as obtain results on  the subdominant and 
phenomenologically interesting electromagnetic  quadrupole form factors.

\end{document}